\documentclass{article}
\usepackage{ijcai24}

\usepackage{times}
\usepackage{soul}
\usepackage{url}
\usepackage[hidelinks]{hyperref}
\usepackage[utf8]{inputenc}
\usepackage[small]{caption}
\usepackage{graphicx}
\usepackage{amsfonts}
\usepackage{amsmath}
\usepackage{amsthm}
\usepackage{booktabs}
\usepackage{algorithm}
\usepackage{algorithmic}
\usepackage{multirow}
\usepackage{multicol}
\usepackage{subfigure}
\usepackage[switch]{lineno}
\usepackage{colortbl}

\title{VisRec: A Semi-Supervised Approach to Radio Interferometric Data Reconstruction}

\author{
Ruoqi Wang$^1$
\and
Haitao Wang$^2$\and
Qiong Luo$^{1,3}$\and
Feng Wang$^4$\And
Hejun Wu$^2$
\affiliations
$^1$HKUST(GZ)\\
$^2$SYSU\\
$^3$HKUST\\
$^4$GZHU
\emails
rwang280@connect.hkust-gz.edu.cn
}

\begin{document}

\maketitle

\begin{abstract}
Radio telescopes produce visibility data about celestial objects, but these data are sparse and noisy. As a result, images created on raw visibility data are of low quality. Recent studies have used deep learning models to reconstruct visibility data to get cleaner images. However, these methods rely on a substantial amount of labeled training data, which requires significant labeling effort from radio astronomers. Addressing this challenge, we propose VisRec, a model-agnostic semi-supervised learning approach to the reconstruction of visibility data. Specifically, VisRec consists of both a supervised learning module and an unsupervised learning module. In the supervised learning module, we introduce a set of data augmentation functions to produce diverse training examples. In comparison, the unsupervised learning module in VisRec augments unlabeled data and uses reconstructions from non-augmented visibility data as pseudo-labels for training. This hybrid approach allows VisRec to effectively leverage both labeled and unlabeled data. This way, VisRec performs well even when labeled data is scarce. Our evaluation results show that VisRec outperforms all baseline methods in reconstruction quality, robustness against common observation perturbation, and generalizability to different telescope configurations.
\end{abstract}

\section{Introduction}
In radio astronomy, a radio interferometer consists of an array of antennas, all observing the same area of sky, forming a single, unified telescope. The \emph{visibility} refers to the signal obtained by cross-correlating pairs of antennas, which is in the frequency domain. Visibility data are then used to construct \emph{images} of the celestial objects being observed \cite{dagli2023astroformer}. Due to the incompleteness and noise in the visibility data, the resultant images, known as \emph{dirty images}, are often dominated by artifacts \cite{schmidt2022deep}. As a result, radio interferometric data must be reconstructed before being used in scientific analysis. In this paper, we introduce a semi-supervised method to reconstruct the visibility.

\begin{figure}
\centerline{\includegraphics[height=3.5cm]{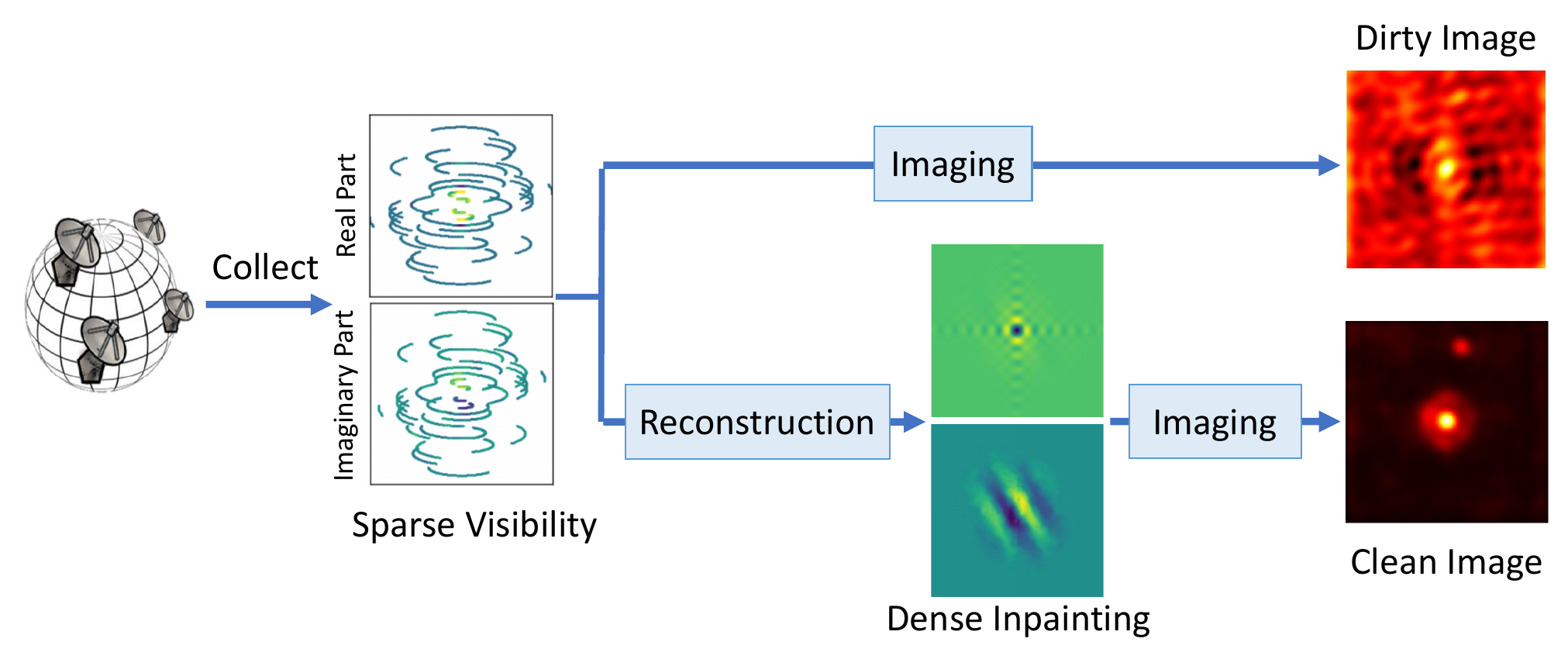}}
\caption{Illustration of radio interferometric data processing. The telescopes collect visibility data. The imaging results of the raw data is dominated by artifacts, called dirty images. In contrast, the imaging results of sparse-to-dense reconstructed visibility data are cleaner.}
\label{process}
\end{figure}

Recently, many deep-learning methods have shown promising reconstruction results on synthetic datasets \cite{sun2021deep,connor2022deep,wu2022neural,schmidt2022deep,WangC0W23,geyer2023deep}. These methods address radio interferometric data reconstruction as a supervised learning problem in which the reconstruction error between the predicted results and the reference labels is minimized \cite{jia2022robustness}. However, these deep-learning-based methods are highly dependent on a great amount of ground truth (labels) for model training, and obtaining ground truth data is difficult. In contrast to natural or medical image reconstruction, where long-time scanning can produce fully-sampled (labeled) training data, obtaining a fully-sampled sky measurement dataset via radio telescopes is impractical due to observational constraints. Furthermore, it takes expertise and human labor to get a ground truth clean image. As such, techniques that can leverage information from unlabeled datasets are indispensable where ground-truth images are scarce. 

Another problem that often comes with the scarcity of labeled data is limited robustness and generalizability in supervised-learning-based reconstruction methods, due to the lack of diversity in training data \cite{darestani2021measuring,desai2023noise2recon}. In particular, existing radio interferometry data reconstructing methods focus on enhancing reconstruction accuracy, and are sensitive to data distribution shift. We followed previous studies \cite{schmidt2022deep,darestani2021measuring} to generate test data with errors caused by noise perturbations during observation or offline antennas as well as data collected from different telescope configurations. Our results confirm the low robustness and generalizability of existing methods.

To address these problems, we propose a model-agnostic semi-supervised learning approach to radio interferometry data reconstruction, called VisRec. Specifically, VisRec consists of two modules: a supervised learning module and an unsupervised learning module. The supervised module integrates a variety of data augmentation functions, exposing the model to diverse training examples. The unsupervised module applies augmentations to unlabeled visibility data and employs reconstructions from non-augmented data as pseudo-labels. VisRec computes the consistency loss between the reconstruction of augmented visibilities and the pseudo-labels to lead the model to be more robust against perturbations. Our method effectively leverages both labeled and unlabeled data to improve model performance.

The main contributions of our work are:
\begin{itemize}
    \item We propose VisRec, which is the first model-agnostic semi-supervised framework for radio interferometric data reconstruction. By combining supervised and unsupervised learning modules, VisRec successfully leverages both labeled and unlabeled data, reducing the dependency on ground truth labels.
    \item We empirically show that our method is robust to common observation perturbation, and has good generalizability when applied to data from different telescope configurations.
    \item We introduce a variety of data augmentation methods for visibility data, including both label-invariant and label-variant augmentations. These augmentations improve the model performance when applied to VisRec as well as supervised learning methods.
\end{itemize}

\section{Background and Related Work}
\subsection{Radio Interferometric Imaging}
In radio astronomy, observing distant astronomical objects via radio interferometry demands large-aperture telescopes, as the angular resolution is inversely related to aperture size \cite{bouman2018reconstructing}. High angular resolution can be obtained using the technique of Very Long Baseline Interferometry (VLBI). VLBI uses a network of globally distributed radio telescopes to create a unified Earth-size telescope. Individual telescopes record radio waves from space, and these signals are then cross-correlated between antenna pairs to produce visibility data. The resulting visibility data is inherently sparse due to the limited number of antennas in VLBI \cite{thompson2017interferometry,bouman2018reconstructing}. 

Visibility data is in the form of complex values in the uv-plane, a geometric plane defined for interferometric observations. The imaging process is an inverse Fourier transformation, translating $(u, v)$ coordinates in the frequency domain into $(l, m)$ coordinates in the image domain \cite{wu2022neural}, represented by the equation:
\begin{eqnarray}
I(l, m) & = & \int_{u} \int_{v} e^{2 \pi i(u l+v m)} V(u, v) d u d v.
\end{eqnarray}
Here, $V(u, v)$ denotes visibility data in the frequency domain, and $I(l, m)$ is the intensity distribution of the sky in the image domain.

\subsection{Interferometric Data Reconstruction}
The imaging results of the under-sampled sparse visibility data are dominated by artifacts so the data must be reconstructed before being used in scientific research \cite{schmidt2022deep}. Interferometric image reconstruction is an ill-posed problem because there are infinite solution images that fit the observed visibility \cite{sun2021deep}. Traditional methods, particularly the CLEAN method \cite{hogbom1974aperture}, have been widely used for this task. CLEAN iteratively refines dirty images to differentiate actual celestial structures from artifacts. However, the performance of CLEAN is limited due to its assumption of point-like sources and long inference time \cite{connor2022deep}.
Recently, some deep-learning-based data processing methods
have been proposed for radio interferometric data reconstruction. Some of them first transfer the sparse visibility data into dirty images through the imaging process and then reconstruct the dirty images into \emph{clean images} \cite{bouman2016computational,sun2021deep,connor2022deep}. In contrast, some recent deep learning-based studies \cite{schmidt2022deep,wu2022neural,geyer2023deep} have proposed to first do inpainting on the visibility data to reconstruct the sparse samples into dense coverage, and then perform imaging to obtain the clean image, showing better performance. However, existing deep-learning methods for interferometric data reconstruction are based on supervised learning, which requires a great amount of training data and does not leverage unlabeled measurements. In this paper, we adopt the reconstruction-and-imaging processing flow and propose a semi-supervised method to leverage both labeled and unlabeled data.


\section{Our Method}

\subsection{Problem Formulation}
In radio astronomy, the task of recovering the brightness distribution of the real sky from the observed visibility data is an inverse problem \cite{sun2021deep}. Consider an image $x \in \mathbb{R}^{n}$ representing the brightness distribution of the sky, and its corresponding sampled visibility $v \in \mathbb{R}^{m}$. They follow the forward model:
\begin{eqnarray}
v=\mathbf{P}_{\Omega} F(x)+\widetilde{\epsilon}.
\end{eqnarray}
In this model, $F$ denotes the Fourier transformation, $\mathbf{P}_{\Omega} \in \mathbb{R}^{m \times n}$ is an under-sampling matrix where $\Omega$ is the telescope's sampling pattern, and $\widetilde{\epsilon}$ denotes noise with the same dimension as $v$. In this paper, our goal is to first estimate the full visibility data $F(x)$ from the observed sparse visibility data, and then apply inverse Fourier transform $F^{-1}$ to estimate the final clean image $x$ of the brightness distribution of real sky.

Consider a dataset $\mathcal{D} = \mathcal{D}^{(s)} \cup \mathcal{D}^{(u)}$. $\mathcal{D}^{(s)}$ denotes the subset consisting of sparse visibility samples as well as corresponding ground truth images for supervised learning. $v^{(s)}_i \in \mathcal{D}^{(s)}$ is the sparse visibility measurement of the $i^{th}$ example in $\mathcal{D}^{(s)}$. $x^{(s)}_i \in \mathcal{D}^{(s)}$ is the image ground truth corresponding to $v^{(s)}_i$.  $\mathcal{D}^{(u)}$ is the subset including sparse visibility data lacking ground truth labels for unsupervised learning. $v^{(u)}_j \in \mathcal{D}^{(u)}$ is the $j^{th}$ example of sparse visibility measurement in $\mathcal{D}^{(u)}$. The reconstruction model, denoted by \( f_\theta \), is parameterized by \( \theta \). \( |\cdot| \) denotes the dataset size. In radio astronomical observation, the size of the supervised dataset \( |\mathcal{D}^{(s)}| \) is smaller than the unsupervised dataset \( |\mathcal{D}^{(u)}| \).

\begin{figure}
\centerline{\includegraphics[height=6.4cm]{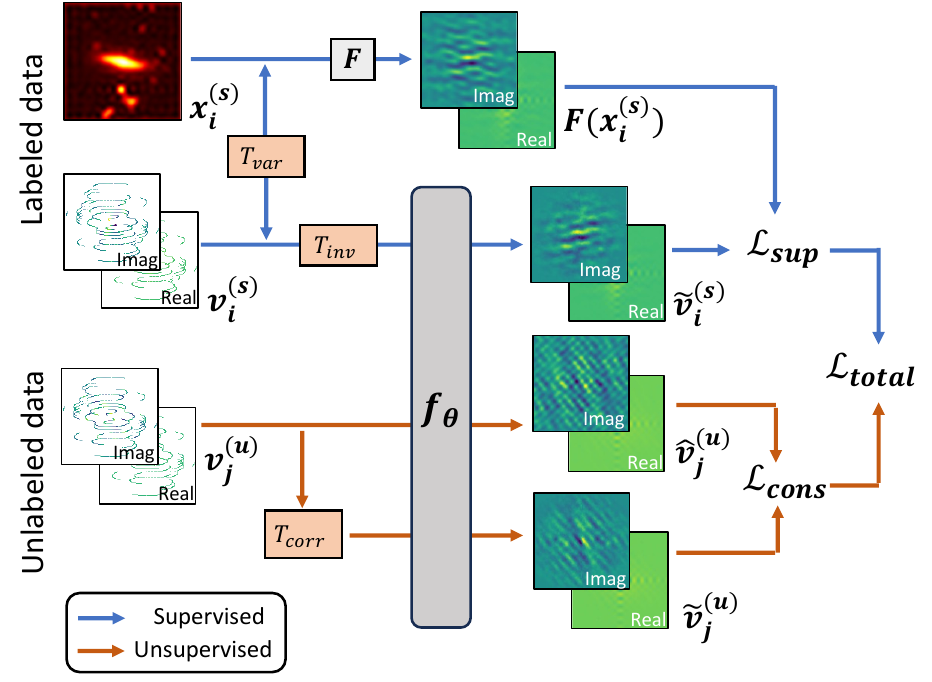}}
\caption{Overview of our method. In our semi-supervised framework, labeled data undergo supervised training, as depicted by the blue arrows. The labeled data are augmented with label-variant and label-invariant augmentations $T_{var}$ and $T_{inv}$. Then the neural network, denoted by $f_{\theta}$, processes augmented visibility data to produce reconstructions. These reconstructions are then compared against ground-truth references to compute supervised loss $\mathcal{L}_{sup}$. Unlabeled sparse visibility data are augmented by $T_{corr}$, and the same $f_{\theta}$ reconstructs both the non-augmented and augmented visibility data. The reconstruction from the non-augmented unlabeled data serves as a pseudo-label for that with augmentations to compute the consistency loss $\mathcal{L}_{cons}$. The overall loss combines supervised and consistency losses in a weighted sum: $\mathcal{L}_{total} \gets \mathcal{L}_{sup} + \lambda  \mathcal{L}_{cons}$.}
\label{overview}
\end{figure}

\subsection{Label-Efficient Reconstruction of Visibility}
We introduce VisRec, a semi-supervised learning framework that employs an arbitrary learnable reconstruction model $f_{\theta}$, parameterized by weights $\theta$, to reconstruct the visibility data into dense inpaintings. Then the dense inpaintings are used to produce high-fidelity clean images. VisRec consists of two modules, a supervised module and an unsupervised module.

\subsubsection{Supervised Module}
In supervised radio interferometry data reconstruction, learning-based models are trained with datasets where reference clean images are available, denoted as $\mathcal{D}^{(s)}$. For each instance, the ground-truth clean image $x$ is transformed into the reference visibility data $F(x)$ to be compared with sparse visibility $v_i^{(s)}$. Here, $F$ is the transformation from the image domain to the visibility domain. The training objective of the model $f_{\theta}$ is to minimize the supervised loss function:

\begin{equation}
\min_{\theta} \frac{1}{|\mathcal{D}^{(s)}|} \sum_{i=0}^{|\mathcal{D}^{(s)}|} \mathcal{L}_{sup} \left(f_{\theta} \left(v_i^{(s)}\right), F\left(x_i^{(s)}\right)\right)
\end{equation}
where $\mathcal{L}_{sup}$ represents the supervised loss function. The model $f_{\theta}$ refers to any parameterized learnable model.

To improve data diversity in label-scarce scenarios, supervised reconstruction techniques can employ data augmentation strategies. We define a series of label-invariant augmentation functions ${T_{inv,l}}, {l=1} ... {m}$ and label-variant augmentation functions ${T_{var,k}}, {k=1} ... {n}$, each applied to the visibility data with a certain probability:

\begin{equation}
T_{inv}(v_i^{(s)}) = T_{inv,1}^{(p_1)} \circ \ldots \circ T_{inv,m}^{(p_m)}(v_i^{(s)})
\end{equation}
\begin{equation}
T_{var}(v_i^{(s)}, x_i^{(s)}) = T_{var,1}^{(q_1)} \circ \ldots \circ T_{var,n}^{(q_n)}(v_i^{(s)},  x_i^{(s)})
\end{equation}
where $p_j, q_k$ are the probabilities of applying each augmentation function.

The augmented loss for an instance with label-invariant visibility augmentation is:
\begin{equation}
\mathcal{L}_{sup} \left(f_{\theta} \left(T_{inv}\left(v_i^{(s)}\right)\right), F\left(x_i^{(s)}\right)\right).
\end{equation}

The augmented loss for an instance with label-variant visibility augmentation is:
\begin{equation}
\mathcal{L}_{sup} \left(f_{\theta} \left(v_i^{\prime{(s)}}\right), F\left(x_i^{\prime{(s)}}\right)\right) 
\end{equation}
where
\begin{equation}
{v}^{\prime (s)}_i, {x}^{\prime(s)}_i = T_{var}(v^{(s)}_i, x^{(s)}_i).
\end{equation}
This dual augmentation approach ensures that the learnable models are trained with a comprehensive range of variations. Suppose an instance of sparse visibility observation is in the form of $\{u_{s}, v_{s}, {V}\left(u_{s}, v_{s}\right)\}$, where $u_{s}$ and $v_{s}$ are coordinates at which a measurement is sampled, and ${V}\left(u_{s}, v_{s}\right)$ is the complex value of the sample. For label-invariant visibility augmentation, we introduce the following methods:
\begin{itemize}
    \item Position Offset:
    \begin{equation}
    (u_{s}', v_{s}') = (u_{s}, v_{s}) + \epsilon_{pos}
    \end{equation}
    where $\epsilon_{pos} \in \mathbb{R}^{\text{shape}((u_{s}, v_{s}))} \sim \mathcal{N}(0, \sigma_{\text{pos}})$ is Gaussian noise added to the position coordinates.
    \item Visibility Noise: 
    \begin{equation}
    V'(u_{s}, v_{s}) = V(u_{s}, v_{s}) + \epsilon_{vis}
    \end{equation}
    where $\epsilon_{vis} \in \mathbb{C}^{\text{shape}(V(u_{s}, v_{s}))} \sim \mathcal{N}(0, \sigma_{vis})$ is complex Gaussian noise added to the visibility values.
    \item Global Position Offset: 
    \begin{equation}
    (u_{s}', v_{s}') = (u_{s}, v_{s}) + (\Delta u, \Delta v)
    \end{equation}
    where $(\Delta u, \Delta v)$ is a uniform random offset applied to all position coordinates.
    \item Random Cropping: A random subset of visibility data is dropped out.
    \item Subset of Selective Frequency:
    The selected sampling subset is represented as:
    \[
     \left\{ (u_s, v_s, V(u_s, v_s)) \, \bigg| \, d_{\text{min}} \leq \sqrt{u_s^2 + v_s^2} \leq d_{\text{max}} \right\}
    \]

where $d_{\text{min}}$ and $d_{\text{max}}$ are the minimum and maximum frequencies of the selected range.

\end{itemize}

For label-variant visibility augmentation, we introduce some geometric transformations in the uv-plane. To augment sparsely sampled visibility data $\{u_{s}, v_{s}, {V}\left(u_{s}, v_{s}\right)\}$ while maintaining correspondence with the image ground-truth $x$, the augmentations must be consistent across both the visibility data and the image domain:
\begin{itemize}
    \item Transpose Transformation: Swap $u_{s}$ and $v_{s}$ coordinates for the visibility data while apply a 90-degree rotation to the clean image $x$. 
    \item Reflection: Negate the $u_{s}$ or $v_{s}$ coordinates in the visibility data. For the clean image $x$, apply a horizontal or vertical flip. 
    \item Central Symmetry Transformation: Negate both $u_{s}$ and $v_{s}$ in the visibility data and apply a central inversion to the clean image $x$. 
\end{itemize}

We design these augmentations following traditional image augmentation techniques \cite{xu2023comprehensive} and adapt them to the properties of visibility data.

\subsubsection{Unsupervised Module}
The unsupervised module of VisRec uses unlabeled visibility data $\mathcal{D}^{(u)}$ that miss ground-truth images. Our aim is to leverage these data to train the model $f_{\theta}$ to recover the original signal from a corruption-augmented measurement \cite{desai2023noise2recon}. Each unsupervised example $v^{(u)}_i$ is augmented by a predefined signal corruption model $T_{corr}$, which simulates expected corrupted patterns that occur in radio interferometry observations. The corruption model $T_{corr}$ applied to each unsupervised instance simulates a variety of realistic observational impairments:

\begin{itemize}
\item Observation Noise \cite{schmidt2022deep}: Introducing zero-mean Gaussian noise to simulate electronic and atmospheric noise.

\item Missing Data \cite{nasirudin2022characterizing}: Simulating the effect of offline antennas by masking out some visibility points.

\item Antenna Offset \cite{nasirudin2022characterizing}: Introducing perturbations that represent errors in antenna positioning or alignment by adding noise to the positional coordinates.
\end{itemize}

Similar to label-invariant visibility augmentation in the supervised learning module, $T_{corr}$ is also applied to the visibility data with a predefined probability. 

The unsupervised training objective is to minimize the following loss function:

\begin{equation}
\min_{\theta} \frac{1}{|\mathcal{D}^{(u)}|} \sum_{j=0}^{|\mathcal{D}^{(u)}|} \mathcal{L}_{cons} \left(f_{\theta} \left(T_{corr}(v^{(u)}_j)\right), f_{\theta} \left(v^{(u)}_j\right)\right)
\end{equation}

This approach trains the model to reconstruct the corrupted visibility by learning to reverse the effects of the corruption-augmentation, which we refer to as \emph{consistency training}. 
\subsubsection{Semi-Supervised Training} 
VisRec consists of both supervised learning and unsupervised learning processes. The main framework of our method is described in Algorithm \ref{alg:VisRec}. In the supervised process, we sample a batch of $\{ (v^{(s)}_i, x^{(s)}_i) \}_{i=1}^{N_s}$ from $\mathcal{D}^{(s)}$. The visibility data $v^{(s)}_i$ and ground-truth image labels $x^{(s)}_i$ are augmented by label-variant augmentation function $T_{var}$. Then the visibility data is augmented by label-invariant augmentation function $T_{inv}$. Then the model $f_{\theta}$ generates the reconstruction of input visibility data.

In the unsupervised process, we sample the batch $\{ v^{(u)}_j \}_{j=1}^{N_u}$ from $\mathcal{D}^{(u)}$. Then each $v^{(u)}_j$ is augmented by corruption model $T_{corr}$. Finally, the same model $f_{\theta}$ reconstructs both augmented and original visibility.
 
Finally, a supervised loss $\mathcal{L}_{sup}$ is computed using the mean squared error between the Fourier-transformed ground truth images and the reconstructed visibility data for the supervised subset. Meanwhile, a consistency loss $\mathcal{L}_{cons}$ is calculated for the unsupervised data by comparing the model's output on the original and augmented visibility samples. The total loss $\mathcal{L}_{total}$ is a weighted sum of these two losses, where $\lambda$ is a hyperparameter that balances the two components. The model parameters $\theta$ are updated by backpropagation to minimize $\mathcal{L}_{total}$.

\begin{algorithm}[tb]
    \caption{Main process of VisRec.}
    \label{alg:VisRec}
    \textbf{Input}: Dataset $\mathcal{D} = \mathcal{D}^{(s)} \cup \mathcal{D}^{(u)}$\\ 
    \textbf{Output}: Trained model $f_\theta$
    \begin{algorithmic}[1] 
        \FOR{each sampled batch $\{ (v^{(s)}_i, x^{(s)}_i) \}_{i=1}^{N_s}$ and $\{ v^{(u)}_j \}_{j=1}^{N_u}$}
            \FORALL{$i \in \{1, \ldots, N_s\}$}
                \STATE ${v}^{\prime (s)}_i, {x}^{\prime(s)}_i \gets T_{var}(v^{(s)}_i, x^{(s)}_i)$
                \STATE $\tilde{v}^{(s)}_i \gets f_\theta(T_{inv}(v^{\prime (s)}_i))$
            \ENDFOR
            \FORALL{$j \in \{1, \ldots, N_u\}$}
                \STATE $\tilde{v}^{(u)}_j \gets f_\theta(v^{(u)}_j)$
                \STATE $\hat{v}^{(u)}_j \gets f_\theta(T_{corr}({v}^{(u)}_j))$
            \ENDFOR
            \STATE $\mathcal{L}_{sup} \gets \frac{1}{N_s} \sum_{i=1}^{N_s} \mathcal{L}_{sup} (F(x^{\prime (s)}_i), \tilde{v}^{(s)}_i)$
            \STATE $\mathcal{L}_{cons} \gets \frac{\lambda}{N_u} \sum_{j=1}^{N_u} \mathcal{L}_{cons}(\hat{v}^{(u)}_j, \tilde{v}^{(u)}_j)$
            \STATE $\mathcal{L}_{total} \gets \mathcal{L}_{sup} + \lambda  \mathcal{L}_{cons}$
            \STATE Update network $f_\theta$ to minimize $\mathcal{L}_{total}$
        \ENDFOR
        \STATE \textbf{return} $f_\theta$
    \end{algorithmic}
\end{algorithm}

\definecolor{gray}{rgb}{0.92, 0.92, 0.92}
\begin{table*}[t]
\fontsize{7pt}{8.5pt}\selectfont
  \centering
  \caption{Comparison of image reconstruction quality on EHT and VLBA datasets (mean and standard deviation). Rows with gray background denote our method and its variants, and the rows with white background represent methods from prior studies.}
  \begin{tabular}{ l|ccc|ccc }
    \hline
    & \multicolumn{3}{c|}{EHT Dataset} & \multicolumn{3}{c}{VLBA Dataset} \\
    \cline{2-7}
    Method & LFD ($\downarrow$) & SSIM ($\uparrow$) & PSNR (dB) ($\uparrow$) & LFD ($\downarrow$) & SSIM ($\uparrow$) & PSNR (dB) ($\uparrow$) \\
    \hline
    Dirty Image & N/A & 0.685 (0.059) & 10.58 (1.28) & N/A & 0.745 (0.041) & 12.24 (1.34) \\
    CLEAN \cite{hogbom1974aperture} & N/A & 0.830 (0.032) & 19.17 (2.61) & N/A & 0.823 (0.031) & 18.48 (2.52) \\
    Noise2Astro \cite{zhang2022noise2astro} & N/A & 0.704 (0.055) & 11.09 (1.36) & N/A & 0.761 (0.034) & 13.08 (1.37) \\
    \hline
    Neural Field \cite{wu2022neural} & 1.083 (0.328) & 0.875 (0.030) & 21.36 (2.51) & 1.152 (0.312) & 0.874 (0.032) & 20.26 (2.60) \\
    \rowcolor{gray}
    NF Supervised + Aug  & 1.043 (0.343) & 0.883 (0.028) & 21.89 (2.47) & 1.083 (0.334) & 0.888 (0.030) & 21.24 (2.75) \\
    \rowcolor{gray}
    NF Self-Supervised  & 1.465 (0.234) & 0.703 (0.056) & 11.45 (1.45)  & 1.433 (0.239) & 0.766 (0.018) & 13.29 (0.90) \\
    \rowcolor{gray}
    NF VisRec w/o Sup-Aug & 1.029 (0.347) & 0.877 (0.033) & 22.53 (2.73) & 1.070 (0.348) & 0.880 (0.033) & 22.22 (2.83)   \\
    \rowcolor{gray}
    \textbf{NF VisRec}  & \textbf{1.029 (0.337)} &  \textbf{0.886 (0.029)} & \textbf{22.63 (2.64)} &  \textbf{1.067 (0.318)} &  \textbf{0.889 (0.031)} & \textbf{22.87 (3.06)} \\
    \hline

    CNN \cite{schmidt2022deep} & 1.061 (0.237) & 0.864 (0.024) & 22.22 (2.46) & 0.949 (0.446) & 0.875 (0.048) & 22.15 (2.59) \\
    \rowcolor{gray}
    CNN Supervised + Aug & 0.991 (0.387) & 0.870 (0.044) & 22.32 (4.19) & 0.950 (0.446) & 0.875 (0.048) & 23.21 (4.60) \\
    \rowcolor{gray}
    CNN Self-Supervised  & 1.481 (0.226)  & 0.785 (0.020)  & 15.18 (1.52) & 1.369 (0.222) & 0.793 (0.022) & 14.65 (1.32) \\
    \rowcolor{gray}
    CNN VisRec w/o Sup-Aug  & \textbf{0.897 (0.245)} & 0.873 (0.029) & 22.60 (3.20) & 0.847 (0.251) & 0.888 (0.029) & 23.62 (3.13) \\
    \rowcolor{gray}
    \textbf{CNN VisRec} & 0.950 (0.292) & \textbf{0.892 (0.027)} & \textbf{23.37 (2.73)} & \textbf{0.793 (0.255)} & \textbf{0.899 (0.024)} & \textbf{23.94 (2.44)} \\
    
    \hline
  \end{tabular}
  \label{tab:comparison_eht_vlba}
\end{table*}

\vspace{-0.2cm}
\section{Experiments}
In this section, we evaluate how well VisRec (1) perform in comparison with existing state-of-the-art non-learning, supervised, and self-supervised methods in label-scarce scenarios and (2) improves the robustness and generalizability of reconstruction models.

\subsection{Experimental Settings}
\textbf{Datasets}.
Following the latest visibility reconstruction studies \cite{wu2022neural,schmidt2022deep,WangC0W23,geyer2023deep}, we use two telescope configurations to sample the visibility observation from real astronomical images to build the visibility dataset. For reference images, we use the Galaxy10 DECals dataset \cite{Galaxy10} which contains 17,736 images of various galaxies. The dataset is from the DESI Legacy Imaging Surveys \cite{dey2019overview}, which merges data from the Beijing-Arizona Sky Survey (BASS) \cite{zou2017project}, the DECam Legacy Survey (DECaLS) \cite{blum2016decam}, and the Mayall z-band Legacy Survey \cite{silva2016mayall}. Using these images as a reference, we employ the eht-imaging toolkit \cite{chael2019ehtim,chael2018interferometric} to produce visibility data represented by $\left\{u_{s}, v_{s}, {V}\left(u_{s}, v_{s}\right)\right\}$. The parameters for observation are adjusted to mirror an 8-telescope Event Horizon Telescope (EHT) setup \cite{wu2022neural}, with the EHT being one of the most prominent arrays leveraging VLBI techniques. In addition, following \cite{schmidt2022deep,geyer2023deep}, we apply the Very Long Baseline Array (VLBA) uv-coverage to simulate another dataset. In EHT dataset, each image has 1,660 visibility points sampled whereas in VLBA dataset, the number of visibility points is 2,298. The image dimensions are all set at 256 $\times$ 256 pixels. For each dataset, we randomly select 1,024 examples for testing and the remaining samples are for training. More details regarding the datasets are reported in the appendix. To simulate a label-scarce setting, we randomly split a small number of examples from the training set as the labeled dataset, and the remaining samples are all used as unlabeled data.

\textbf{Platform}. 
We conduct all experiments on a server with two AMD EPYC 7763 CPUs, 512GB main memory, and eight Nvidia RTX 4090 GPUs each with 24GB device memory. The server is equipped with two NVME 2TB SSD and two 16TB SATA hard disks. The operating system is Ubuntu 20.04. Our model is implemented in PyTorch 1.8.1 \cite{paszke2019pytorch}.

\textbf{Evaluation Metrics}. 
To evaluate frequency data differences, we use Log Frequency Distance (LFD) \cite{jiang2021focal}, where lower values indicate better performance. To assess quality of recovered images, we apply Peak Signal-to-Noise Ratio (PSNR) and Structural Similarity Index Measure (SSIM), both computed via the scikit-image package \cite{singh2019basics}. PSNR assesses overall image quality, and SSIM evaluates perceptual similarity to ground truth. A higher PSNR and SSIM is better.


\subsection{Baseline Methods}
In this study, we compare our method with a classic non-learning method CLEAN \cite{hogbom1974aperture}, the self-supervised denoising method for astronomical image Noise2Astro \cite{zhang2022noise2astro}, and two state-of-the-art supervised learning methods for visibility reconstruction \cite{wu2022neural,schmidt2022deep}. Radionet \cite{schmidt2022deep} uses a convolutional neural network (CNN) as the reconstruction model and Neural Interferometry \cite{wu2022neural} uses a neural field model (NF). We apply VisRec to these two models, and we also introduce three variants of our method as baseline approaches:
\begin{itemize}
    \item Supervised Learning with Data Augmentation: Adding the proposed augmentation methods to the supervised training of the NF and CNN model, denoted NF Supervised + Aug and CNN Supervised + Aug, respectively. Essentially, this approach represents the VisRec framework with its unsupervised consistency training component removed.
    \item Self-Supervised Training: This is a variant of VisRec that is extended to a fully unsupervised setting. In this setup, the supervised training pathway is replaced with the self-supervised training setup from Noise2Noise \cite{lehtinen2018noise2noise}. 
    \item VisRec without Augmentation in Supervised Module: In this variant of VisRec, we remove all augmentations in the supervised module to evaluate the impact of consistency learning separately, denoted VisRec w/o Sup-Aug. 
\end{itemize}

More implementation details of these methods are included in the appendix.

\begin{figure}[t]
\centerline{\includegraphics[height=3cm]{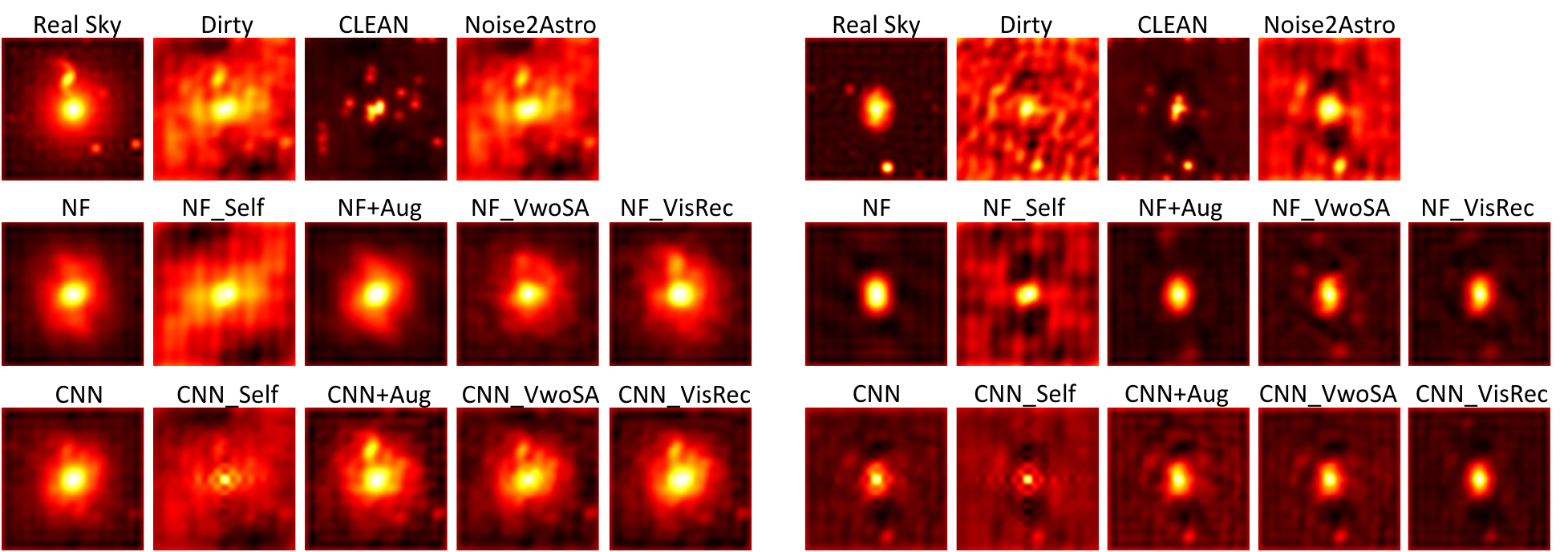}}
\caption{Visual Examples of overall comparison. VwoSA denotes VisRec w/o Sup-Aug.}
\label{overallcomp}
\end{figure}

\subsection{Overall Comparison}
\label{Overall Comparison}
To assess the performance of our method in scenarios with limited label availability, we conduct an overall comparison using only 1,024 labeled visibility examples (\( |\mathcal{D}^{(s)}| = 1,024\)) and a set of 15,668 unlabeled data (\( |\mathcal{D}^{(u)}| \) = 15,668). We calculate the LFD, PSNR, and SSIM values for all test data reconstructed using our method and other baseline methods, presenting both mean values and standard deviations in Table \ref{tab:comparison_eht_vlba}. The results show that both NF model \cite{wu2022neural} and CNN model \cite{schmidt2022deep} trained by supervised methods outperform traditional non-learning method (CLEAN) and self-supervised Noise2Astro \cite{zhang2022noise2astro}. Our VisRec approach, applied to both neural field and CNN models, shows significant improvements in all metrics compared to the original supervised methods. Its variant model Supervised + Aug and VisRec w/o Sup-Aug also shows better reconstruction quality than the supervised versions, demonstrating that both the augmentation and the consistency training in VisRec are effective. We also present two sets of representative reconstructed images from EHT dataset in Figure \ref{overallcomp}.

\begin{figure}[b]
\centerline{\includegraphics[height=5.4cm]{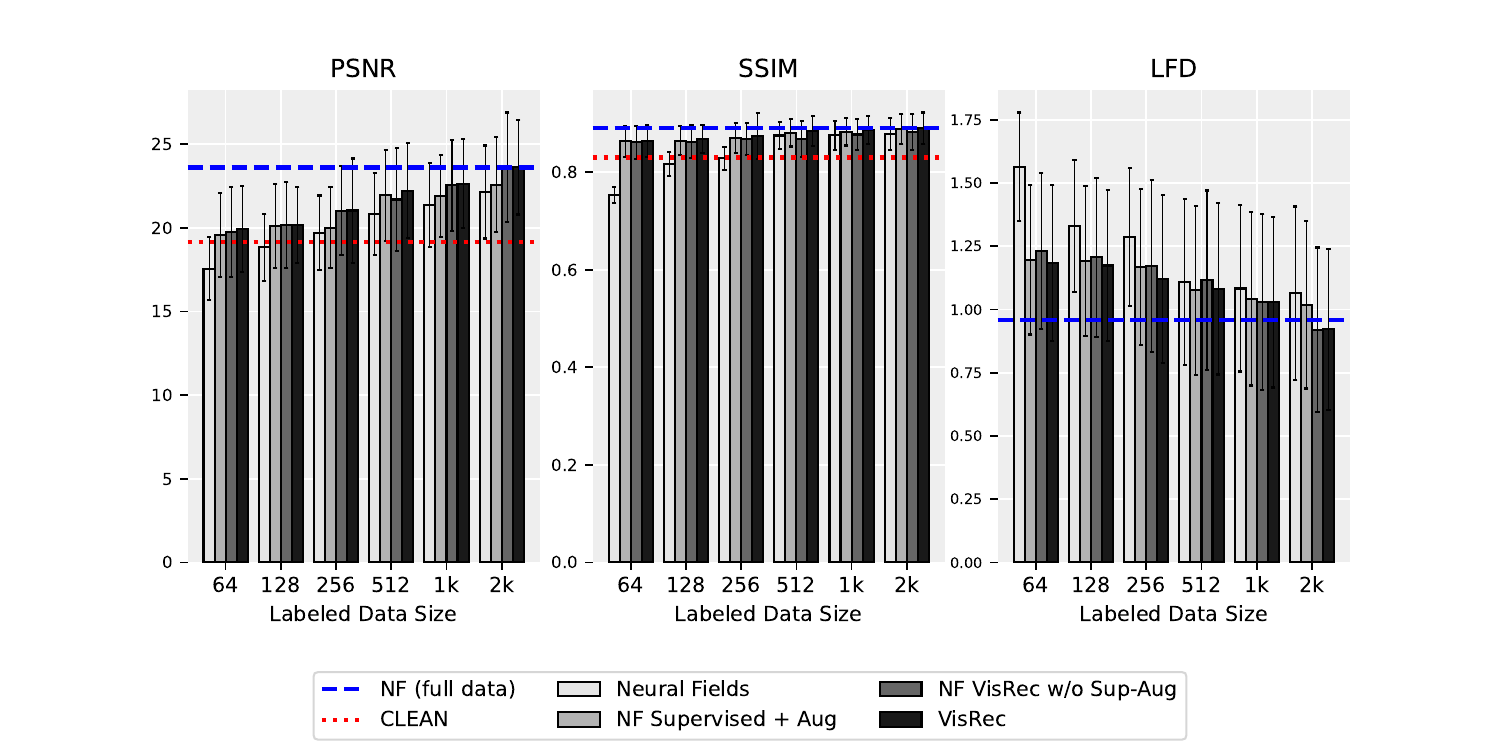}}
\caption{Effect of Labeled Data Size.}
\label{data_size}
\end{figure}

\subsection{Effect of Labeled Data Size}
\label{Effect of Labeled Data Size}
We also evaluate the performance of our method and other baselines under various labeled data sizes. We apply VisRec, along with its two variants (Supervised + Aug and VisRec w/o Sup-Aug) to train the NF model \cite{wu2022neural} on the EHT dataset. Maintaining a constant total training data size \( |\mathcal{D}| \) = 16,692, we vary the size of the labeled subset \( |\mathcal{D}^{(s)}| \) from 64 to 2k (k = 1024) and use the remaining data as the unlabeled subset. We also evaluate the performance of the original supervised model trained with all 16,692 labeled data examples, denoted NF (Full data). As shown in Figure \ref{data_size}, our method and its variants consistently outperform the original supervised method. With only 64 labeled examples, our method outperforms the traditional non-learning method CLEAN. With only 2k labeled examples, VisRec performs on par with supervised methods trained with over 16k examples. The results show that the effectiveness of VisRec comes from the use of both supervised and unsupervised data, making it well-suited for limited label scenarios. Moreover, incorporating data augmentation also improves model performance by enhancing training diversity.

In addition, we report the effect of labeled data size on VisRec performance with the CNN model in the appendix.

\subsection{Robustness Evaluation}
\label{Robustness Test}

In this experiment, we evaluate the robustness of our method and the baselines under the same dataset setting as in section \ref{Overall Comparison}.

\begin{figure}[t]
\centerline{\includegraphics[height=4.7cm]{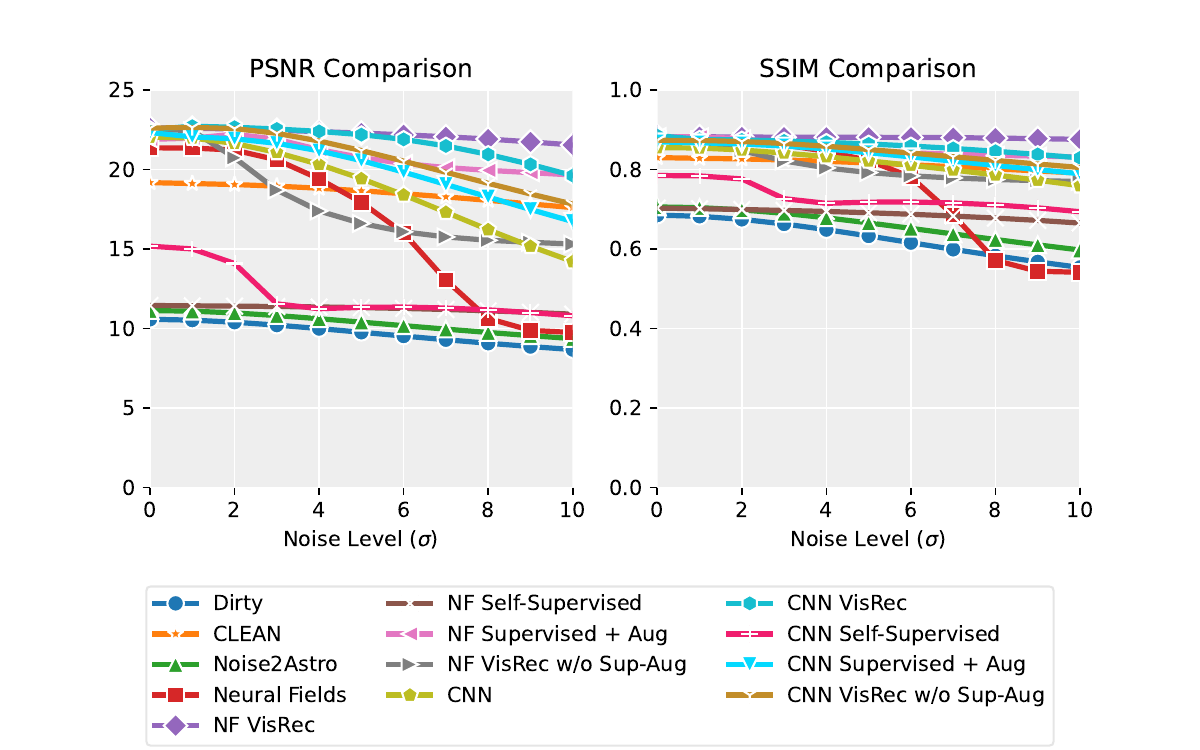}}
\caption{Performance of different methods across various noise levels. We only show PSNR and SSIM values because LFD is not applicable to Dirty, CLEAN, and Noise2Astro.}
\label{noise}
\end{figure}

\textbf{Robustness against Noise.} In radio astronomy, both instrumental and environmental factors can introduce noise \cite{schmidt2022deep}, which may significantly degrade the quality of the images reconstructed by previous supervised learning methods. To evaluate the robustness of various methods to such noise, we conduct performance tests across different noise levels. 

\begin{figure}[ht]
\centerline{\includegraphics[height=2.9cm]{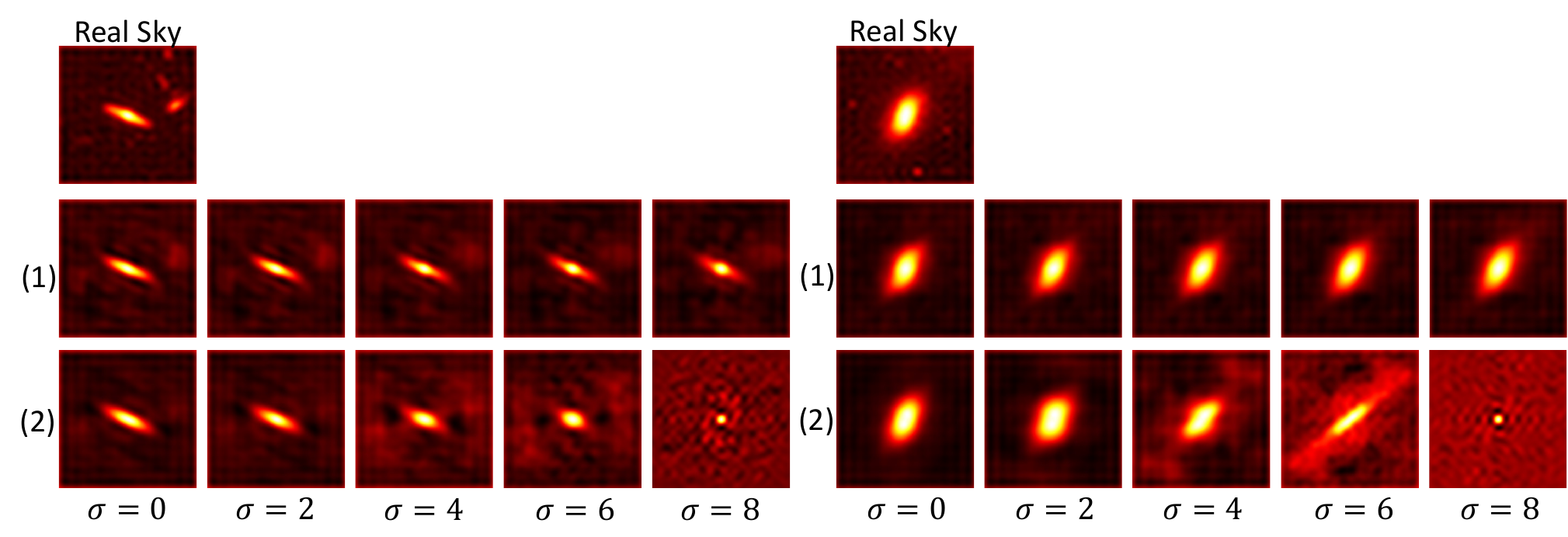}}
\caption{Visual examples of the effect of noise. Row (1) is the results of the NF model trained by VisRec and row (2) is the results of original supervised NF method. The examples are all from the EHT dataset.}
\label{eg_noise}
\end{figure}

Following Schmidt et al. \cite{schmidt2022deep}, we add white noise that simulates potential measurement effects in the frequency domain by corrupting the visibilities with the noise of Gaussian distribution. This process can be represented as:

\begin{equation}
V_{\text{noisy}}(u_s, v_s) = V(u_s, v_s) + g(x | \mu, \sigma)_{uv}
\end{equation}

\noindent where $g(x | \mu, \sigma)_{uv}$ is a random value generated for each frequency sample, with $\mu = 0$ and noise level $\sigma = 0 \text{ to } 10$. Figure \ref{noise} shows that models applied VisRec consistently outperform other methods and have lower sensitivity to noise. The two variant of our method (Supervised + Aug and VisRec w/o Sup-Aug) also show better robustness then the original supervised methods. We also display some visual examples in Figure \ref{eg_noise}, showing that as the noise level increases, the reconstruction of supervised learning methods are severely corrupted, whereas the reconstruction results of VisRec still maintain high quality.

\begin{figure}[h]
\centerline{\includegraphics[height=4.7cm]{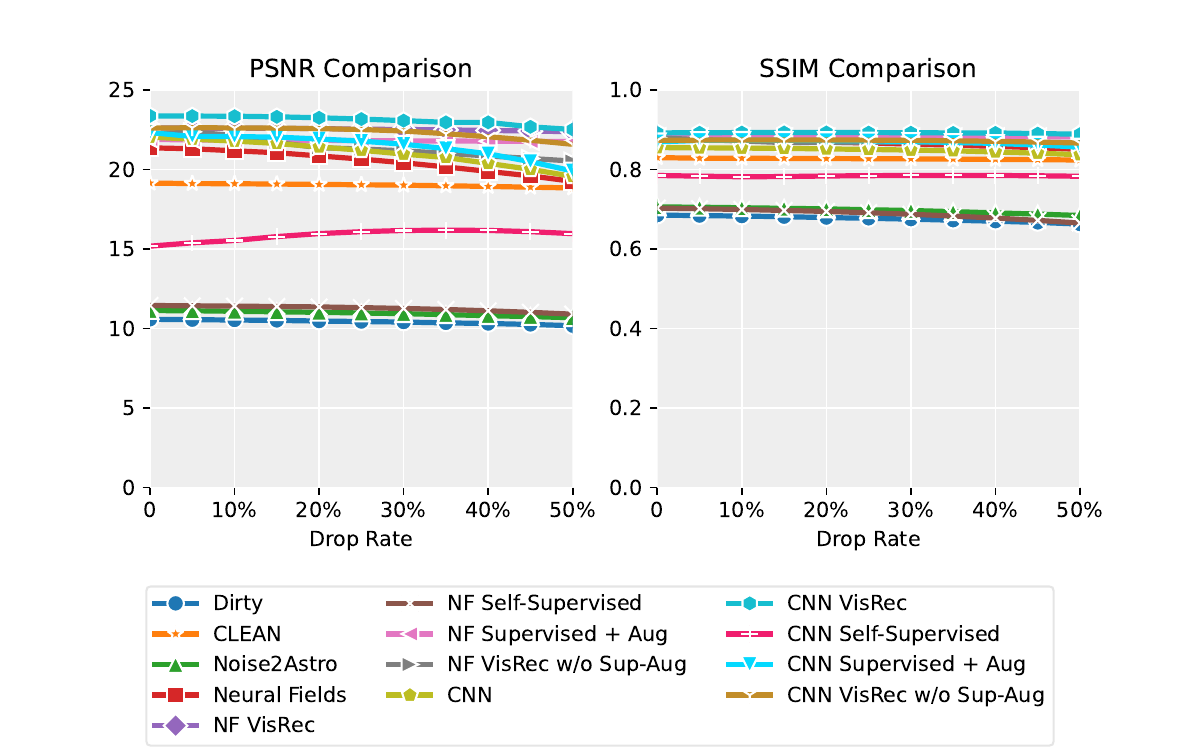}}
\caption{Performance of different methods across various sample loss rates. We only show PSNR and SSIM values because LFD is not applicable to Dirty, CLEAN, and Noise2Astro.}
\label{drop}
\end{figure}

\textbf{Robustness against Missing Visibility Points.}
Missing visibility points, often resulting from offline antennas in interferometric arrays \cite{nasirudin2022characterizing}, also cause performance decrease of learning-based reconstruction methods. We vary the sample loss rate from 0 to 50\% and evaluate the performance of our method and the baseline methods. Figure \ref{drop} shows that VisRec achieves the best overall sample loss rate and is less sensitive to increasing of missing samples than supervised methods. To show the result more clearly, we list the value of the NF model and CNN model trained with VisRec and the supervised method respectively in Table \ref{drop_table}. The rate of VisRec performance decrease is significantly lower than supervised learning when applied to both NF and CNN models, indicating that VisRec improves the robustness of deep models.

\begin{table}[ht]
\centering
\fontsize{6pt}{8pt}\selectfont
\caption{PSNR/SSIM values of methods at 0 and 50\% sample loss rates.}
\begin{tabular}{c|c|c|c|c}
\hline
\textbf{Loss Rate} & NF Sup & NF VisRec & CNN Sup & CNN VisRec \\
\hline
0 & 21.36 / 0.8754 & 22.63 / 0.8855 & 22.22 / 0.8636 & 23.37 / 0.8920 \\
50\% & 19.25 / 0.8506 & 22.40 / 0.8848 & 19.59 / 0.8372 & 22.51 / 0.8889 \\
\hline
\textit{Decrease} & -9.88\% / -2.83\% & -1.02\% / -0.08\% & -11.84\% / -3.06\% & -3.68\% / -0.35\% \\
\hline
\end{tabular}
\label{drop_table}
\end{table}

\vspace{-0.5cm}

\begin{table}[h]
\centering
\tiny
\caption{Performance of training methods with the same or different datasets. Gray rows denote our method and its variants, and the rows in white background represent methods from prior studies.}
\label{tab:Generalizability}
\begin{tabular}{ l|ccc|ccc }
\hline
\multirow{2}{*}{Model} & \multicolumn{3}{c|}{Same Dataset} & \multicolumn{3}{c}{Different Dataset} \\
 & LFD & SSIM & PSNR & LFD & SSIM & PSNR  \\
\hline
Neural Field \cite{wu2022neural} & 1.083  & 0.875  & 21.36 & 1.319 & 0.819 & 16.74 \\
\rowcolor{gray}
NF Supervised + Aug & 1.043  & 0.883  & 21.89  & 1.305 & 0.842 & \textbf{18.95} \\
\rowcolor{gray}
NF Self-Supervised & 1.465 & 0.703 & 11.45 & 1.489 & 0.595 & 9.55 \\
\rowcolor{gray}
NF VisRec w/o Sup-Aug & 1.029 & 0.877 & 22.53  & \textbf{1.242} & 0.844 & 18.42 \\
\rowcolor{gray}
\textbf{NF VisRec} & \textbf{1.029} & \textbf{0.886} & \textbf{22.63}  & 1.250 & \textbf{0.855} & 18.65 \\
\hline
CNN \cite{wu2022neural} & 1.061  & 0.864  & 22.22 & 1.141 & 0.802 & 16.13  \\
\rowcolor{gray}
CNN Supervised + Aug & 0.991  & 0.870 & 22.32 & 1.238 & 0.823 & 18.28  \\
\rowcolor{gray}
CNN Self-Supervised & 1.481  & 0.785  & 15.17 & 1.409 & 0.736 & 11.88  \\
\rowcolor{gray}
CNN VisRec w/o Sup-Aug & \textbf{0.897} & 0.873 & 22.60 & 1.044 & 0.822 & 16.99  \\
\rowcolor{gray}
\textbf{CNN VisRec} & 0.950  & \textbf{0.892}  & \textbf{23.37}  & \textbf{1.001} & \textbf{0.866} & \textbf{20.65}  \\
\hline
\end{tabular}
\end{table}

\vspace{-0.3cm}

\subsection{Generalization to Another Telescope}
We evaluate the generalizability of different methods by training the models on the EHT dataset (the same training setting as in Section \ref{Overall Comparison}) and testing their performance on both EHT and VLBA datasets. 
As shown in Table \ref{tab:Generalizability}, the significant performance drop in all models when applied to different datasets underscores the challenge in model generalization. VisRec, VisRec w/o Sup-Aug, and Supervised+Aug methods outperform the traditional supervised method, especially on different datasets, indicating superior generalizability. While in some cases the variants of our method, e.g., VisRec w/o Sup-Aug, and Supervised+Aug show marginally better performance than VisRec in certain metrics, the overall performance of VisRec is superior. Particularly, the CNN model trained with VisRec shows the least performance degradation, suggesting its high generalizability in handling diverse data distributions.

\subsection{Impact of Consistency Loss Weight}
\begin{figure}[h]
\centerline{\includegraphics[height=4cm]{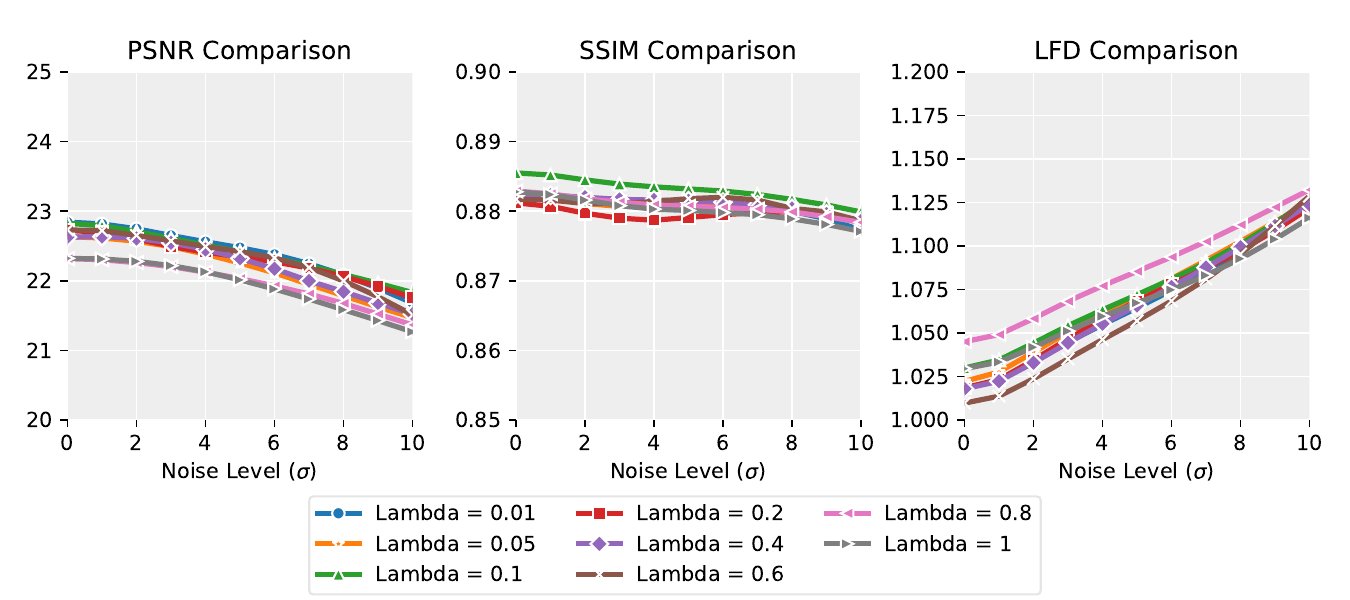}}
\caption{Impact of $\lambda$. The differences are small so we have narrowed the range of the y-axis for close observation.}
\label{lambda}
\end{figure}

We evaluate the impact of consistency loss weight $\lambda$ in VisRec on the reconstruction of visibilities with different noise corruption levels. The results are shown in Figure \ref{lambda}. The performance of VisRec does not change much for a large range of $\lambda \in \left[0.01, 0.8\right)$. Insensitivity to changes in $\lambda$ may help eliminate the need for hyperparameter tuning, which can simplify network training \cite{desai2023noise2recon}.

\section{Conclusion}
In this study, we present VisRec, a semi-supervised learning framework for radio interferometric data reconstruction. VisRec effectively leverages both labeled and unlabeled data, significantly reducing dependence on extensive ground truth labels. Our results show that VisRec outperforms state-of-the-art methods in reconstruction quality, robustness to common observational perturbations, and generalizability across different telescope configurations.




\clearpage

\bibliographystyle{named}
\bibliography{ijcai24}

\end{document}